\documentclass [12pt]{article}
\usepackage{amsmath,amssymb,cite}

\usepackage{feynmp}
\usepackage[vcentermath,enableskew]{youngtab}
\usepackage{tabularx}
\usepackage[english]{babel}
\usepackage[dvips]{graphicx}
\usepackage{indentfirst}
\usepackage{epsfig}
\numberwithin{equation}{section}

\begin{document}

\title{On the Problem of Vacuum Energy in FLRW Universes and Dark Energy}

\author{Badis Ydri \footnote{Email:ydri@stp.dias.ie,~badis.ydri@univ-annaba.org.}, Adel Bouchareb  \footnote{Email:adel.bouchareb@univ-annaba.org.}\\
%\email{}
%\affiliation{ 
Institute of Physics, BM Annaba University, BP 12, 23000, Annaba, Algeria.
}
%\date{}
\maketitle
\begin{abstract}
We present a (hopefully) novel calculation of the vacuum energy in expanding FLRW spacetimes based on the renormalization of quantum field theory  in non-zero backgrounds. We compute the renormalized effective action up to the $2-$point function and then apply the formalism to the cosmological backgrounds of interest. As an example we calculate for quasi de Sitter spacetimes the leading correction to the vacuum energy given by the tadpole diagram  and show that it behaves as    $\sim H_0^2\Lambda_{\rm pl}$ where $H_0$ is the Hubble constant and $\Lambda_{\rm pl}$ is the Planck constant. This is of the same order of magnitude as the observed dark energy density in the universe.   
\end{abstract}

\maketitle

\section{Introduction}
In recent years it has been established, to a very reasonable level of confidence, that the universe is spatially flat and is composed of $4$ per cent ordinary mater, $23$ per cent dark matter and $73$ per cent dark energy. This state of affair is summarized in the cosmological  concordance $\Lambda$CDM model  \cite{Beringer:1900zz}. %See also the pedagogical presentations \cite{Carroll:2000fy, Baumann:2009ds} for a much simpler discussion  and for more references to the original literature.  
The dominant component, dark energy, is believed to be the same thing as the cosomlogical constant introduced by Einstein in $1917$ which in turn is believed to originate in the energy of the vacuum \cite{Weinberg:1988cp}. We note that dark energy is characterized mainly by a negative pressure and no dependence on the scale factor and its density behaves as $\sim H_0^2\Lambda_{\rm pl}$ where $H_0$ is the Hubble parameter and $\Lambda_{\rm pl}=1/\sqrt{8\pi G}$ is the Planck mass \cite{Carroll:2000fy}. More precisely we have (with $\Omega_{\Lambda}\simeq 0.73$) \footnote{The vacuum energy density is a constant which can be expressed as something which is proportional to the square of the Hubble parameter at the current epoch (Hubble constant). We are not saying that the vacuum energy density  falls with the square of the Hubble parameter.}
\begin{eqnarray}
\rho_{\Lambda}=3\Omega_{\Lambda}H_0^2\Lambda_{\rm pl}^2
&\simeq &39\Omega_{\Lambda}(10^{-12}{\rm GeV})^4.\label{observation}
\end{eqnarray}
The reality of the energy of the vacuum is exhibited, as is well known, in a dramatic way in the Casimir force. See for example \cite{milton} for a systematic presentation of this subject. 

In this article we will adopt the usual line of thought outlined in \cite{Weinberg:1988cp,Carroll:2000fy} and identify dark energy with vacuum energy.

The calculation of vacuum energy in curved spacetimes such as FLRW universes and de Sitter spacetime requires the use of  quantum field theory in the presence of a non zero gravitational background \cite{Birrell:1982ix,Fulling:1989nb}. de Sitter spacetime is of particular interest since we know that both the early universe as well as its future evolution is dominated by vacuum, i.e. FLRW universes may be understood as a perturbation $V$ of de Sitter spacetime which vanishes in the early universe as well as in the limit $a\longrightarrow \infty$. The main difficulties in doing quantum field theory on curved background is the definition of the vacuum state and renormalization. 

In an expanding de Sitter spacetime and also in quasi de Sitter spacetimes a natural choice of the vacuum is given by the so-called Euclidean or Bunch-Davies state \cite{Bunch:1978yq,Mottola:1984ar,Allen:1985ux}. It can be shown \cite{angus,inprogress} that the vacuum energy in de Sitter spacetime with the Bunch-Davies state behaves in the right way as $\sim H_0^2\Lambda_0^2$ where $H_0$ is the de Sitter Hubble parameter and $\Lambda_0$ is a physical cutoff introduced for example along the lines of \cite{Kempf:2000ac}. As discussed above FLRW spacetimes may be treated as quasi de Sitter spacetimes. The usual in-out formalism may then be used to extend  the calculation of vacuum energy to these spacetimes \cite{inprogress}.

A more systematic and more fundamental approach to the calculation of vacuum energy in FLRW spacetimes is based on the renormalization of quantum field theory  in non-zero backgrounds. This is the approach we will discuss in this article which is inspired by the recent original work on the Casimir force found in \cite{Jaffe:2005vp,Graham:2003ib,Milton:2002vm} in which the starting point is a renormalizable quantum field theory in a non-zero background. The main ingredients of this approach are as follows:
\begin{itemize}
\item[1)]We reinterpret scalar field theory coupled to FLRW metric as a scalar field theory in a flat spacetime interacting with a particular time-dependent (effective graviton) background. 
\item[2)]We regularize and then renormalize the resulting  scalar field theory for arbitrary backgrounds in the usual way.  Renormalization, if possible, removes all divergences from all proper vertices of the effective action. 
\item[3)]We compute the vacuum energy for arbitrary backgrounds.
\item[4)]We substitute the particular background found in $1)$. 
\item[5)]There is always the possibility that the vacuum energy still diverges for particular profiles of the background configuration. This is indeed the case for the Casimir force  \cite{Graham:2003ib} as well as for the FLRW spacetimes considered here. We thus regularize with a cutoff to obtain an estimate for the vacuum energy.
\end{itemize}
Although we think that this approach is novel, potential and possible overlap with other approaches is certainly expected. A systematic investigation of this point is still underway.
\section{The Main Result}
In the following we will be interested in spatially flat  Friedmann-Lema\^itre-Robertson-Walker (FLRW) universes with line elements given by

\begin{eqnarray}
ds^2&=&-dt^2+a^2(t)d\vec{x}^2\nonumber\\
&=&a^2(\eta)(-d\eta^2+d\vec{x}^2).
\end{eqnarray}
The $x_i$ are the comoving coordinates, $\eta$ is the conformal time and $a$ is the scale factor. The action of a real massless scalar field coupled to this metric is given by
\begin{eqnarray}
S_{\phi}&=& \int d^4x\sqrt{-{\rm det}g}~\bigg(-\frac{1}{2}g^{\mu\nu}\nabla_{\mu}\phi\nabla_{\nu}\phi-\frac{1}{2}\xi R\phi^2\bigg)\nonumber\\
&=&\int d^4x  \frac{a^2}{2}~\bigg(\phi^{'2}-(\partial_i\phi)^2-\xi a^2 R\phi^2\bigg).
\end{eqnarray}
The scalar curvature is given by $R=6a^{''}/a$ where the prime stands for the derivation with respect to $\eta$. The path integral and the expectation values of the theory are
\begin{eqnarray}
{\cal O}_{\phi}=\frac{1}{Z}\int {\cal D}\phi~{\cal O}(\phi)~e^{iS_{\phi}}~,~Z=\int {\cal D}\phi~e^{iS_{\phi}}.
\end{eqnarray}
We perform the change of variable $\phi\longrightarrow \chi =a\phi$ where $\chi$ is the comoving field. The action $S_{\phi}$ becomes

\begin{eqnarray}
S_{\chi}
&=&\int d^4x  \frac{1}{2}~\bigg(\chi^{'2}+\frac{a^{''}}{a}\chi^2-(\partial_i\chi)^2-\xi a^2 R\chi^2\bigg).\label{actionchi}
\end{eqnarray}
The path integral and the expectation values of the theory  in terms of $\chi$ are
\begin{eqnarray}
{\cal O}_{\chi}=\frac{1}{Z}\int {\cal D}\chi~{\cal O}(\chi)~e^{iS_{\chi}}~,~Z=\int {\cal D}\chi~e^{iS_{\chi}}.
\end{eqnarray}
In general ${\cal O}_{\phi}\neq {\cal O}_{\chi}/a$. We can check for example that the Hamiltonians in terms of $\phi$ and $\chi$, which are defined using the stress-energy-momentum tensor in the usual way, are related by 
\begin{eqnarray}
{H}_{\phi}=\frac{1}{a}{H}_{\chi}-\frac{1}{2a}\bigg[\frac{a^{''}}{a}M_{\chi}+\partial_{\eta}(\frac{a^{'}}{a}M_{\chi})\bigg].\label{fundamental}
\end{eqnarray}
The second moment is defined by
\begin{eqnarray}
M_{\chi}=<\int d^3x\chi^2>.
\end{eqnarray}
The goal now is to compute $H_{\chi}$ and $M_{\chi}$. This requires the quantization of the scalar field $\chi$ in a time dependent (effective graviton) background 
\begin{eqnarray}
\sigma=(\xi-1/6)a^2R.\label{sub}
\end{eqnarray}
 The action is of course given by  (\ref{actionchi}) which can also be  rewritten including a mass $M$ for the field $\chi$ as 

\begin{eqnarray}
S_{\chi}
&=&\int d^4x  ~\bigg(-\frac{1}{2}\partial_{\mu}\chi\partial^{\mu}\chi-\frac{1}{2}M^2\chi^2-\frac{1}{2}\sigma\chi^2\bigg).\label{actionS}
\end{eqnarray}
In the following we will first assume an arbitrary background and then substitute the particular background (\ref{sub}) at the end.  It is natural to expect that UV divergences arise in the quantum theory which warrants therefore regularization and renormalization. The theory given by the action (\ref{actionS}) is renormalizable. In fact it can be used to construct an inductive proof for the renormalizability of phi-four theory  \cite{ZinnJustin:2002ru}. We observe that only the $1-$point and the $2-$point functions are superficially divergent in the theory given by the action (\ref{actionS}). All higher order correlation functions are finite. We will use dimensional continuation as a regulator. Despite the fact that the energy is time dependent in our case as opposed to the Casimir energy a counter term of the form ${\cal L}_{\rm ct}=\delta_1\sigma+\delta_2\sigma^2$ along the lines of \cite{Jaffe:2005vp,Graham:2003ib} is  sufficient to remove all divergences. Thus in order to renormalize the theory we add the counterterm action 
\begin{eqnarray}
S_{\rm ct}=\int d^4x (\delta_1\sigma+\delta_2\sigma^2).\label{ct}
\end{eqnarray}
In the spirit of normal ordering instead of computing $H_{\chi}[\sigma]$ and $M_{\chi}[\sigma]$ we compute the shift in these expectation values with the respect to the case with zero background \cite{Fulling:1989nb}. We have then
\begin{eqnarray}
{\cal H}_{\chi}[\sigma]&=&H_{\chi}[\sigma]-H_{\chi}[0]\nonumber\\
&=&\frac{1}{2}\int d^3x~\big(\partial_0^x\partial_0^y-(\partial_0^y)^2\big)<x|\big(D_{\sigma}-D_{0}\big)|y>.\nonumber\\
\end{eqnarray}
\begin{eqnarray}
{\cal M}_{\chi}[\sigma]&=&M_{\chi}[\sigma]-M_{\chi}[0]\nonumber\\
&=&-2\int d^3x \bigg[\frac{\delta W}{\delta\sigma(x)}[\sigma]-\frac{\delta W_{}}{\delta\sigma(x)}[0]\bigg].
\end{eqnarray}
The two-point function $D_{\sigma}$ is the inverse of the Laplacian $i(\partial_{\mu}\partial^{\mu}-M^2-\sigma)$ while $W[\sigma]$ is the vacuum energy (the generating energy functional of all connected Green's functions)  given by $W=-i \ln Z$. 

\begin{figure}[htbp]
\begin{center}
\includegraphics[width=8cm,angle=0]{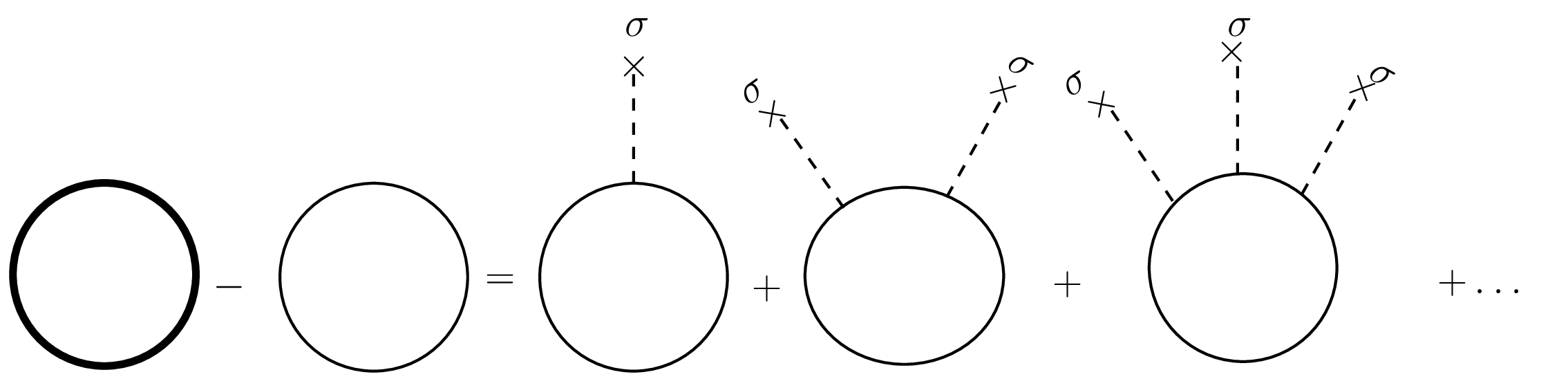}
\caption{The Feynman diagrams contributing to the vacuum energy.
}
\label{graph}
\end{center}
\end{figure}

Generically ${\cal H}_{\chi}$ and ${\cal M}_{\chi}$ are time-dependent. Let  $\tilde{\cal H}_{\chi p}$ and $\tilde{\cal M}_{\chi p}$ be the Fourier transforms of ${\cal H}_{\chi}$ and ${\cal M}_{\chi}$ respectively defined by (with $p=(p^0,0,0,0)$)
\begin{eqnarray}
\tilde{\cal H}_{{\chi} p}=\int d\eta {\cal H}_{\chi}~e^{-ip x}~,~\tilde{\cal M}_{{\chi} p}=\int d\eta {\cal M}_{\chi}~e^{-ip x}.
\end{eqnarray}
The diagrammatic expansion of ${\cal H}_{\chi}$ and ${\cal M}_{\chi}$ is given by the sum of all one-loop Feynman diagrams with at least one external leg $\sigma$. See Figure (\ref{graph}). Both the $1-$ and $2-$point functions contributions to    ${\cal H}_{\chi}$ are divergent at $d=4$ while only the $1-$point function contribution to ${\cal M}_{\chi}$ is divergent at $d=4$. We note that the  $1-$point function contribution to    ${\cal H}_{\chi}$  is also divergent for $d<4$.

In order to subtract the divergences arising in this theory we need to tune the counterterms $\delta_1$ and $\delta_2$ introduced (\ref{ct}) appropriately. In our case here we will employ a combination of modified minimal subtraction and renormalization conditions. We introduce a mass scale $\mu^2$ and require that the $1-$point proper vertex $\Gamma^{(1)}(p)=<\tilde{\sigma}(p)>_{1{\rm PI}}$ vanishes at this scale, viz  
\begin{eqnarray}
\Gamma^{(1)}(p)|_{p^2=-\mu^2}=0.
\end{eqnarray}
We also require that the $2-$point proper vertex $\Gamma^{(2)}(p,k_2)=<\tilde{\sigma}(-k_2+p)\tilde{\sigma}(k_2)>_{1{\rm PI}}$ vanishes at $p=0$ and $k_2^2=-\mu^2$, viz  
\begin{eqnarray}
\Gamma^{(2)}(p,k_2)|_{p=0,k_2^2=-\mu^2}=0.
\end{eqnarray}
In the case of a time independent background configuration this reduces to the usual condition $\Gamma^{(2)}(k_2)|_{k_2^2=-\mu^2}=0$. The mass scale $\mu^2$ in the above two equations may not be necessarily the same. 

By computing the $1-$point and the $2-$point functions, taking into account the counterterms and imposing the  above renormalization conditions we obtain after some calculation the values of the counterterms. We find explicitly
\begin{eqnarray}
\delta_1&=&\frac{M^2}{32\pi^2}\bigg(\frac{2}{\epsilon}+1-\gamma+\ln 4\pi\bigg)-\frac{1}{32\pi^2}\bigg[\frac{1}{6}\mu^2+\int ds  \big(M^2+s(3s-2)\mu^2\big)\ln \big(M^2-s(1-s)\mu^2\big)\bigg].\label{CT1}\nonumber\\
\end{eqnarray}
\begin{eqnarray}
\delta_2
&=&-\frac{1}{64\pi^2}\bigg(\frac{2}{\epsilon}-\gamma+\ln 4\pi \bigg)+\frac{1}{32\pi^2} \int_0^1 ds \bigg[(1-s)\ln \big(M^2-s(1-s)\mu^2\big)+ \frac{2s(s-1)\mu^2}{M^2-s(1-s)\mu^2}\bigg].\label{CT2}\nonumber\\
\end{eqnarray}
We observe that the $1-$point function (tadpole) contribution to ${\cal H}_{\chi}$ vanishes identically in the limit $M^2\longrightarrow 0$ for any value of $\mu^2$. We end up with the result 
\begin{eqnarray}
\tilde{\cal H}_{\chi p}[\sigma]
&=&\frac{1}{32\pi^2} \int_0^1 ds_1\int_0^{1-s_1}ds_2\int_{k_2}\bigg[\ln \frac{\Delta}{s_2(s_2-1)\mu^2}+\frac{\Delta^{'}}{\Delta}\bigg]\tilde{\sigma}(-k_2+p)\tilde{\sigma}(k_2)+O(\sigma^3).\label{x1x2x3}\nonumber\\
\end{eqnarray}
The second important observation is that the same counterterm $\delta_2$ given by (\ref{CT2}) is sufficient to cancel the divergence arising in the $1-$point function contribution to ${\cal M}_{\chi}$. We find 
\begin{eqnarray}
\tilde{\cal M}_{\chi p}[\sigma]&=&\bigg[\frac{1}{16\pi^2}\ln\frac{(p^0)^2}{\mu^2}-\frac{1}{4\pi^2}\bigg]\tilde{\sigma}(p)\nonumber\\
&-&\frac{1}{32\pi^2} \int_0^1 ds_1\int_0^{1-s_1} ds_2\int_{k_2}\frac{1}{\Delta}\tilde{\sigma}(-k_2+p)\tilde{\sigma}(k_2).
\end{eqnarray}
In the above equations $\Delta$ and $\Delta^{'}$ are defined by
\begin{eqnarray}
\Delta=s_1p^2+s_2k_2^2-(s_1p+s_2k_2)^2.
\end{eqnarray}
\begin{eqnarray}
\Delta^{'}=2(s_1p^0+s_2k_2^0)^2-3p^0(s_1p^0+s_2k_2^0)+(p^0)^2.
\end{eqnarray}
\section{Phenomenological Application}
As a phenomenological application let us use the above formulas to estimate the vacuum energy density in de Sitter spacetime and in FLRW spacetimes thought of as perturbed de Sitter spacetime. For simplicity we will only use the leading correction given by the tadpole diagram and as a consequence only ${\cal M}_{\chi}$ contributes to the vacuum energy ${\cal H}_{\phi}$. 
  
In the remainder we will take the scale factor to be given by the ansatz 
\begin{eqnarray}
a=-\frac{e^{V}}{H_0\eta}.\label{scale}
\end{eqnarray}
For $V=0$ we obtain precisely  de Sitter spacetime. This is the unperturbed solution. We will think of the FLRW universe given by the scale factor (\ref{scale}) as a perturbation of de Sitter spacetime in the following sense. In all inflationary models the early  exponential expansion of the universe corresponds to a spacetime which is very close to de Sitter spacetime. In the language of the $S-$matrix the "in" solution is therefore de Sitter spacetime in the infinite past $\eta\longrightarrow -\infty$. On the other hand observations of distant supernovae indicate that the universe is currently undergoing accelerated expansion driven by a small positive cosmological constant. By excluding the possibility of recollapse we can argue that vacuum energy will dominate over matter in the limit $a\longrightarrow \infty$ and thus FLRW universe will approach de Sitter spacetime in this limit \cite{Carroll:2004st}. The  "out" solution is therefore also de Sitter spacetime in the infinite future $\eta\longrightarrow 0$. The interaction potential which connects between the "in" and "out" solutions is essentially given by the function $V$ in equation (\ref{scale}).

%We expect the existence of two phase transitions. The first is the one associated with the end of inflation and reheating after which matter dominated the universe. In the language of the $S-$matrix this happens when the "in" solution scatters off (hits) the potential $V$. The second phase transition occurs when vacuum energy starts to dominate over matter, i.e. when the "out" solution emerges out of the potential $V$. 

The vacuum energy (\ref{fundamental}) with the scale factor (\ref{scale}) takes now the form 

\begin{eqnarray}
{\cal H}_{\phi}[\sigma]&=&-\frac{1}{a}\big[\frac{a^{''}}{a}-2\frac{a^{'2}}{a^2}\big]{\cal M}_{\chi}-\frac{a^{'}}{2a^2}{\cal P}_{\chi}\nonumber\\
%&=&-\frac{1}{a}\big[V^{''}+\frac{2}{\eta}V^{'}-(V^{'})^2\big]{\cal M}_{\chi}+\frac{1}{2a}(\frac{1}{\eta}-V^{'}){\cal P}_{\chi}\nonumber\\
&=&-\frac{1}{a}\big[2V^{''}-{\cal V}\big]{\cal M}_{\chi}+\frac{1}{2a}(\frac{1}{\eta}-V^{'}){\cal P}_{\chi}.
\end{eqnarray}
The potential ${\cal V}$ is given by
\begin{eqnarray}
{\cal V}=V^{''}-\frac{2}{\eta}V^{'}+(V^{'})^2.
\end{eqnarray}
${\cal M}_{\chi}$ and ${\cal P}_{\chi}$ are given explicitly by (by omitting also the subscript $0$ on the momentum)
\begin{eqnarray}
{\cal M}_{\chi}
&=&\frac{1}{8\pi^2}\frac{v}{a^3}(1-6\xi)\int dp p\big(\ln\frac{(p)^2}{\mu^2}-4\big)\cos p\eta\nonumber\\
&-&\frac{1}{16\pi^2}\frac{v}{a^3}(1-6\xi)\int \frac{dp}{2\pi}\tilde{\cal V}(p)\big(\ln\frac{(p)^2}{\mu^2}-4\big)e^{-ip\eta}.\nonumber\\
\end{eqnarray}
\begin{eqnarray}
{\cal P}_{\chi}
&=&-\frac{1}{8\pi^2}\frac{v}{a^3}(1-6\xi)\int dp (p)^2\big(\ln\frac{(p)^2}{\mu^2}-4\big)\sin p\eta\nonumber\\
&+&\frac{i}{16\pi^2}\frac{v}{a^3}(1-6\xi)\int \frac{dp}{2\pi}p\tilde{\cal V}(p)\big(\ln\frac{(p)^2}{\mu^2}-4\big)e^{-ip\eta}.\nonumber\\
\end{eqnarray}
In the above equation $v$ is the physical volume of spacetime which stems from our use of box normalization and $\tilde{\cal V}(p)$ is the Fourier transform of the potential ${\cal V}(\eta)$. For de Sitter spacetime the vacuum energy reduces to 
%For $V=0$ we have 
\begin{eqnarray}
{\cal H}_{\phi}
%&=&\frac{1}{2a}\frac{1}{\eta}{\cal H}_3\nonumber\\
&=&-\frac{1}{16\pi^2 \eta}\frac{v}{a^4}(1-6\xi)\int_0^{\infty} dp~ p^2\bigg(\ln\frac{p^2}{\mu^2}-4\bigg)\sin p \eta.\nonumber\\
\end{eqnarray}
This expression is well defined in the infrared limit $p\longrightarrow 0$ but divergent in the ultraviolet limit $p\longrightarrow \infty$. This can be traced to the fact that the Fourier transform of the effective graviton configuration given by $a^{''}/a$  does not vanish sufficiently fast for large momenta. Let us then introduce a hard comoving cutoff $\Lambda$. We are then interested in the integral 
\begin{eqnarray}
\int_0^{\Lambda} dp~ p^2\bigg(\ln\frac{p^2}{\mu^2}-4\bigg)\sin p \eta.
\end{eqnarray}
%We use the result obtainable after several operations of integration by parts
%\begin{eqnarray}
%\int_0^{\Lambda} dp~ p^2\bigg(\ln\frac{p^2}{\mu^2}-4\bigg)\sin p \eta %&=&-\frac{1}{\eta}\bigg[2\Lambda^2\ln\Lambda -4\Lambda^2-\Lambda^2\ln\mu^2\bigg]\cos\Lambda\eta+\frac{1}{\eta^2}\bigg[4\Lambda\ln\Lambda -6\Lambda-2\Lambda\ln\mu^2\bigg]\sin\Lambda\eta\nonumber\\
%&-&\frac{1}{\eta^3}\bigg[4\ln\Lambda -2-2\ln\mu^2\bigg]\bigg[1-\cos\Lambda\eta\bigg]+\frac{4}{\eta^3}\int_0^{\Lambda}\frac{dp}{p}[1-\cos p\eta\big]\nonumber\\
%&=&-\frac{1}{\eta}\bigg[2\Lambda^2\ln\Lambda -4\Lambda^2-\Lambda^2\ln\mu^2\bigg]\cos\Lambda\eta\nonumber\\
%&+&\frac{1}{\eta^2}\bigg[4\Lambda\ln\Lambda -6\Lambda-2\Lambda\ln\mu^2\bigg]\sin\Lambda\eta\nonumber\\
%&-&\frac{1}{\eta^3}\bigg[4\ln\Lambda -2-2\ln\mu^2\bigg]\bigg[1-\cos\Lambda\eta\bigg]+\frac{4}{\eta^3}(C+\ln|\eta|\Lambda)\nonumber\\
%&+&\frac{4}{\eta^3}\int_{\Lambda|\eta|}^{\infty}dp\frac{\cos p}{p}.\label{idt}
%\end{eqnarray}
The comoving cutoff $\Lambda$ is related to the physical cutoff $\Lambda_0$ by $\Lambda=a\Lambda_0$ \cite{Kempf:2000ac}. Since on de Sitter $a=-1/(H_0\eta)$ we have $\Lambda=\Lambda_0/(|\eta|H_0)$ and $|\eta|\Lambda=\Lambda_0/H_0$. The limit of interest $\Lambda\longrightarrow \infty$ may then be achieved by letting  $\Lambda_0/H_0\longrightarrow \infty$. 

We take the value of the Hubble parameter of de Sitter spacetime to be the value of the Hubble parameter of the universe at the current epoch, viz
\begin{eqnarray}
H_0\simeq 14.91\times 10^{-43} {\rm GeV}.\label{par1}
\end{eqnarray}
Furthermore by assuming that quantum field theory calculations are valid up to the Planck scale $M_{\rm pl}=1/\sqrt{8\pi G}$ we can take \cite{Weinberg:1988cp}
\begin{eqnarray}
\Lambda_0=M_{\rm pl}\simeq 2.42\times 10^{18} {\rm GeV}.\label{par2}
\end{eqnarray}
It is obvious that with these parameters we have 
\begin{eqnarray}
\frac{\Lambda_0}{H_0}>>1.
\end{eqnarray}
We can then approximate the above integral by \footnote{We remark that although $\Lambda_0/H_0 >>1$ the cosine remains bounded and therefore the remaining divergence, which is due to the special cosmological shapes, is really quadratic.}
\begin{eqnarray}
\int_0^{\Lambda} dp~ p^2\bigg(\ln\frac{p^2}{\mu^2}-4\bigg)\sin p \eta &\simeq &\eta a^4H_0^2\Lambda_0^2\bigg[4-\ln\frac{\Lambda^2}{\mu^2}\bigg]\cos\frac{\Lambda_0}{H_0}.
\end{eqnarray}
The energy density becomes then
\begin{eqnarray}
\frac{{\cal H}_{\phi}}{v}
&=&\frac{1}{16\pi^2 }(1-6\xi)H_0^2\Lambda_0^2\bigg[\ln\frac{\Lambda^2}{\mu^2} -4\bigg]\cos\frac{\Lambda_0}{H_0}.
\end{eqnarray} 
The mass scale $\mu^2$ is also comoving and therefore the physical mass scale  must be defined by $\mu^2=\mu_0^2a$. The energy density is then time independent given by
\begin{eqnarray}
\frac{{\cal H}_{\phi}}{v}
&=&(1-6\xi)\frac{H_0^2\Lambda_0^2}{16\pi^2 }\bigg[\ln\frac{\Lambda_0^2}{\mu_0^2} -4\bigg]\cos\frac{\Lambda_0}{H_0}.
\end{eqnarray} 
The mass scale $\mu_0^2$ may be taken to be of the order of particle physics mass scale, say
\begin{eqnarray}
\mu_0\simeq 10^2{\rm GeV}.\label{par3}
\end{eqnarray} 
By using the parameters (\ref{par1}), (\ref{par2}) and (\ref{par3}) we obtain a numerical estimation for the vacuum energy density given by 
\begin{eqnarray}
\frac{{\cal H}_{\phi}}{v}
&=&(1-6\xi)\big(0.08\times \big(10^{-12}{\rm GeV}\big)^4\big)(71.45)(0.94)\nonumber\\
&=&(1-6\xi)\big(5.37\times \big(10^{-12}{\rm GeV}\big)^4\big).
\end{eqnarray}
This is of the same order of magnitude as the experimental value (\ref{observation}). Corrections due to deviation from a perfect de Sitter spacetime are of order $V$ while corrections due to the contribution of the $2-$point function are of the order of $(1-6\xi)^2$. A quantitative discussion of these effects will be discussed elsewhere \cite{inprogress}. 
 
\section{Conclusion}
In this article we have presented a new calculation of the vacuum energy in a certain class of FLRW spacetimes which can be viewed as perturbed de Sitter spacetime. This calculation is based on the renormalization of quantum field theory in non-zero (effective graviton) backgrounds in analogy with  the recent treatment of the Casimir force found in \cite{Jaffe:2005vp,Graham:2003ib}. It is found that the vacuum energy still diverges, after renormalization of the quantum field theory proper vertices, for the time-dependent cosmological backgrounds of interest. Indeed these backgrounds do not vanish sufficiently fast for large momenta. By introducing an appropriate comoving cutoff \footnote{A comoving cutoff breaks Lorentz invariance but we are only here trying to obtain a rough, albeit reasonable, estimation of the vacuum energy density.} an estimation of the vacuum energy is obtained which is compared favorably with the experimental value. 
 
\paragraph{Acknowledgments:} This research was supported by “The National Agency for the
Development of University Research (ANDRU)” under PNR contract number U23/Av58
(8/u23/2723).

\end{document}